\documentclass[a4paper,8pt]{article}
\usepackage[english]{babel}
\usepackage{graphicx,amsmath,upgreek,float,mathtools}
\usepackage{fancyhdr,listings}
\usepackage{color} 
\definecolor{mygreen}{RGB}{28,172,0} 
\definecolor{mylilas}{RGB}{170,55,241}
\newcommand{\beq}{\begin{equation}}
\newcommand{\eeq}{\end{equation}}

\newcommand{\TITLE}{Practical algorithms for simulation and reconstruction of digital in-line holograms}
\begin{document}
\begin{center}
{\LARGE\bf\TITLE}
\vspace{0.5cm}

Tatiana Latychevskaia and Hans-Werner Fink\\
Physics Department, University of Zurich, Winterthurerstrasse 190, 8057 Zurich, Switzerland\\
Corresponding author: tatiana@physik.uzh.ch

\end{center}
\begin{abstract}
\noindent Here we present practical methods for simulation and reconstruction of in-line digital holograms recorded with plane and spherical waves. The algorithms described here are applicable to holographic imaging of an object exhibiting absorption as well as phase shifting properties. Optimal parameters, related to distances, sampling rate, and other factors for successful simulation and reconstruction of holograms are evaluated and criteria for the achievable resolution are worked out. Moreover, we show that the numerical procedures for the reconstruction of holograms recorded with plane and spherical waves are identical under certain conditions. Experimental examples of holograms and their reconstructions are also discussed.\\

\noindent OCIS: (090.0090) Holography; (090.1995) Digital holography; (090.2880) Holographic interferometry; (110.3010) Image reconstruction techniques; (120.3180) Interferometry; (100.1830) Deconvolution.
\end{abstract}

{\tableofcontents}

\noindent For matlab codes see the link \href{https://ch.mathworks.com/matlabcentral/profile/authors/856133}{\bf matlab file exchange, author: Tatiana Latychevskaia}
\newpage
\section{Introduction}

\noindent In-line holography relates to the original holographic scheme proposed by Gabor \cite{Gabor:1947,Gabor:1948,Gabor:1949}. It is of conceptually simple design, does not include optical elements between sample and detector and has been employed since its invention in numerous experiments using various types of waves, be it light, electrons or X-rays, to name just a few. Nowadays, holograms are recorded by digital detectors and are subject to numerical reconstruction, which constitutes the field of digital holography \cite{Schnars:2005}. A good overview of different types of holograms and the theory dedicated to their formation and reconstruction is given in the book by Kim \cite{Kim:2011}. All simulation and reconstruction routines applied in digital holography employ fast Fourier transforms (FFT). Most of the routines utilize single Fourier transform, except the routine for plane waves based on the angular spectrum method \cite{Goodman:2004}, where two Fourier transforms are involved. In general, optimal reconstructions are achieved when two Fourier transforms are employed \cite{Molony:2010}. The reason is twofold. Firstly, when two Fourier transforms are involved in simulation or reconstruction, the object and its hologram are sampled with a similar number of pixels. For example, if the object occupies a quarter in the object plane, its hologram will also occupy approximately a quarter of the detector area, and vice versa. Secondly, when a single Fourier transform is employed, all of the following parameters are co-dependent and bound by one equation: the distance between sample and detector, the number of pixels, the wavelength, the object area size, and the detector area size. Thus, the correct reconstruction, provided all distances are given by the experimental arrangement, can be achieved only at a certain fixed number of pixels, which is highly inconvenient. On the other hand, a calculation of wave propagation that employs two Fourier transforms makes it possible to avoid such dependency on the number of pixels. Here, we summarize simple methods for simulation and reconstruction of holograms with both plane and spherical waves. All of the algorithms described here employ two Fourier transforms. Some of the methods described here, have been employed in previous studies \cite{Latychevskaia:2007,Latychevskaia:2009,Latychevskaia:2012} but not been discussed in detail.
\section{Hologram Formation and Reconstruction}

\begin{figure}[htbp]
\centerline{\includegraphics[width=12cm]{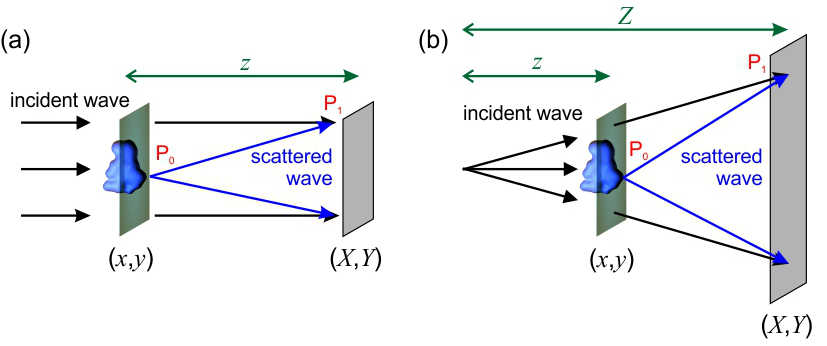}}
\caption{In-line holography schemes realized with (a) a plane wave and (b) a spherical wave.}\label{Fig:1}
\end{figure}

\noindent By definition, in in-line holography, the reference wave and the object wave share the same optical axis. Typically, the experiment is realized as follows: a wave passes by an object located at positions in the  plane. Part of the wave is scattered by the object, thus creating the object wave $O$, and the unscattered part of the wave forms the reference wave $R$. The two waves interfere beyond the object and the interference pattern recorded at some distance is named the hologram. In Fig. \ref{Fig:1} two in-line holography schemes are displayed utilizing plane respectively spherical waves. In-line holography with spherical waves is also called Gabor holography \cite{Gabor:1948,Gabor:1949}.
The incident wave distribution is described by $U_{\rm incident}(x,y)$, with $(x,y)$ being coordinates in the object plane,  $\displaystyle k=\frac{2\pi}{\lambda}$, with $\lambda$ denoting the wavelength. An object is described by a transmission function \cite{Latychevskaia:2007,Latychevskaia:2009}:
\beq\label{Eq:HF01}
t(x,y) =
\exp{[-a(x,y)]}\exp{[i\phi(x,y)]},
\eeq
\noindent where $a(x,y)$ describes the absorption and $\phi(x,y)$ the phase distribution while the wave is scattered off the object. From Eq. (\ref{Eq:HF01}), it is obvious that the transmission function $t(x,y) = 1$ where there is either no object or where $a(x,y) = 0$ and $\phi(x,y) = 0$ implying that the distribution of the incident wave remains undisturbed. This observation allows the object transmission function to be rewritten as:
\beq\label{Eq:HF02}
t(x,y) =
1 + \tilde{t}(x,y),
\eeq
\noindent where $\tilde{t}(x,y)$ is a perturbation imposed onto the reference wave, not necessarily a small term however. Equation \ref{Eq:HF02} is just a mathematical representation to allow for separating contributions from reference respectively object wave. The wavefront distribution beyond the object, the so-called exit wave, is then described by:
\beq\label{Eq:HF03}
U_{\rm exit\; wave}(x,y) =
U_{\rm incident}(x,y) \cdot t(x,y) =
U_{\rm incident}(x,y) + U_{\rm incident}(x,y) \cdot \tilde{t}(x,y),
\eeq
\noindent where the first term describes the reference and the second term describes the object wave.

The propagation of the wave towards the detector is described by the Fresnel-Kirchhoff diffraction formula:
\beq\label{Eq:HF04}
U_{\rm detector}(X,Y) =
-\frac{i}{\lambda}
\int\int U_{\rm incident}(x,y) \cdot t(x,y)
\frac{\exp{\left( ik \left|\vec{r} - \vec{R} \right| \right)}}{\left|\vec{r} - \vec{R} \right|}\;
{\rm d} x {\rm d} y,
\eeq
\noindent where $\left|\vec{r}_{\rm P_0} - \vec{r}_{\rm P_1} \right| = \left|\vec{r} - \vec{R} \right|$ denotes the distance between a point in the object plane $\rm P_0$ and a point in the detector plane $\rm P_1$, as illustrated in Fig. \ref{Fig:1}. Here $\vec{r} = (x,y,z)$ and $\vec{R} = (X,Y,Z)$.

The distribution of the two waves at a detector positioned in the plane $(X,Y)$ is described by $R(X,Y)$ and $O(X,Y)$, respectively. The transmission of the recorded hologram is therefore given by:
\beq\label{Eq:HF05}
H(X,Y) = \left| U_{\rm detector}(X,Y) \right|^2 =
\left| R(X,Y)\right|^2 + \left| O(X,Y)\right|^2 +
+ R^{*}(X,Y)O(X,Y) + R(X,Y)O^{*}(X,Y),
\eeq
\noindent where the first term is the constant background created by the reference wave alone, the second term is assumed to be small compared to the strong reference wave term, and the last two terms give rise to the interference pattern observed in the hologram.

Before reconstruction, the hologram must be normalized by division with the background image:
\beq\label{Eq:HF06}
B(X,Y) =
\left| R(X,Y)\right|^2.
\eeq

\noindent The background image is recorded under the exact same experimental conditions as the hologram, but without the object being present. The distribution of the normalized hologram
\beq\label{Eq:HF07}
H_0(X,Y) =
\frac{H(X,Y)}{B(X,Y)} - 1 \approx
\frac{R^{*}(X,Y)O(X,Y) + R(X,Y)O^{*}(X,Y)}{\left| R(X,Y)\right|^2}
\eeq
\noindent does thus not depend on such factors as the intensity of the incident or reference wave or detector and camera sensitivity. After the normalization procedure, the hologram can be reconstructed by applying routines that are described below. The subtraction of $1$ leaves only the interference term, which approaches $0$ wherever the object wave approaches $0$, for example at the edges of the hologram. Thus, the hologram $H_0(X, Y)$ has a smaller folding-fringe effect on its edges due to Fourier transformation. In addition, an apodization cosine window filter is applied to the hologram to minimize effects due to the edges of the hologram caused by the digital Fourier transform (see Appendix A).

The reconstruction of a digital hologram consists of a multiplication of the hologram with the reference wave $R(X,Y)$ followed by back-propagation to the object plane based on the Fresnel-Kirchhoff diffraction integral:
\beq\label{Eq:HF08}
U(x,y) \approx \frac{i}{\lambda}
\int\int R(X,Y) H_0(X,Y)
\frac{\exp{\left( - ik \left|\vec{r} - \vec{R} \right| \right)}}{\left|\vec{r} - \vec{R} \right|}\;
{\rm d} X {\rm d} Y.
\eeq
\noindent The wavefront reconstructed from $H_0(X, Y)$ corresponds to $\tilde{t}(x,y)$ and $1$ should be added to the reconstruction to obtain the transmission function $t(x,y)$ as follows from Eq. (\ref{Eq:HF02}). Finally, Eq. (\ref{Eq:HF01}) is applied to extract absorption and phase distributions of the imaged object.
\section{In-line Holography with Plane Waves}

In this section we describe methods of simulating and reconstructing holograms created with plane waves, as illustrated in Fig. \ref{Fig:1}(a). A plane wave is described by a complex-valued distribution $\exp{\left( i (k_x x + k_y y + k_z z) \right)}$, where $(k_x, k_y, k_z)$ are the components of the wave vector. By selecting the optical axis along the propagation of the plane wave, we obtain $k_x = k_y = 0$, and by choosing the origin of the $z$-axis so that $z = 0$ at the object location, we obtain the incident wave:
\beq\label{Eq:PW01}
U_{\rm incident} (x,y) = 1.
\eeq
\noindent The exit wave behind the object given by Eq. (\ref{Eq:HF03}) equals:
\beq\label{Eq:PW02}
U_{\rm exit\; wave} (x,y) = t(x,y).
\eeq
\noindent The wave propagating from the object plane $(x, y)$ towards the detector plane $(X, Y)$ is described by the Fresnel-Kirchhoff diffraction formula, see Eq. (\ref{Eq:HF04}):
\beq\label{Eq:PW03}
U_{\rm detector} (X,Y) =
-\frac{i}{\lambda}
\int\int t(x,y) \frac{\exp{\left( ik \left|\vec{r} - \vec{R} \right| \right)}}{\left|\vec{r} - \vec{R} \right|} \;
{\rm d} x {\rm d} y,
\eeq
\noindent where
\beq\label{Eq:PW04}
\left|\vec{r} - \vec{R} \right| =
\sqrt{\left( x - X \right)^2 + \left( y - Y \right)^2 + z^2}.
\eeq
\noindent The reconstruction of a digital hologram recorded with plane waves is given by Eq. (\ref{Eq:HF08}) where $R(X,Y) = 1$:
\beq\label{Eq:PW05}
U(x,y) \approx \frac{i}{\lambda}
\int\int H_0(X,Y)
\frac{\exp{\left( - ik \left|\vec{r} - \vec{R} \right| \right)}}{\left|\vec{r} - \vec{R} \right|}\;
{\rm d} X {\rm d} Y.
\eeq
\subsection{Large $z$-distance, Fresnel Approximation}

When the $z$ distance is sufficiently large so that the Fresnel approximation
\beq\label{Eq:PWFA01}
z^3 \gg \frac{\pi}{4\lambda}
\left[ \left( x - X \right)^2 + \left( y - Y \right)^2 \right]_{\rm max}^2
\eeq
\noindent is fulfilled, Eq. (\ref{Eq:PW03}) turns into:
\beq\label{Eq:PWFA02}
U_{\rm detector}(X,Y) =
- \frac{i}{\lambda z}
\int\int t(x,y)
\exp{\Biggl\{
\frac{i\pi}{\lambda z}\left[\left( x - X \right)^2 + \left( y - Y \right)^2 \right]
\Biggr\}} \;
{\rm d} x {\rm d} y,
\eeq
\noindent where the constant phase factor was neglected. Equation (\ref{Eq:PWFA02}) can be re-written in the form of a convolution
\beq\label{Eq:PWFA03}
U_{\rm detector}(X,Y) =
t(X,Y) \otimes s(X,Y)
\eeq
\noindent of the object transmission function $t(x,y)$ with the Fresnel function:
\beq\label{Eq:PWFA04}
s(x,y) =
- \frac{i}{\lambda z} \exp{\left[ \frac{i\pi}{\lambda z} \left( x^2 + y^2 \right) \right]},
\eeq
\noindent whose Fourier transform $S(u,v)$ is given by:
\beq\label{Eq:PWFA05}
S(u,v) =
- \frac{i}{\lambda z} \int\int
\exp{\left[ \frac{i\pi}{\lambda z} \left(x^2 + y^2 \right) \right]}
\exp{\left( - 2 \pi i \left( x u + y v \right) \right)} \;
{\rm d} x {\rm d} y =
\exp{\left[ - i \pi \lambda z \left( u^2 + v^2 \right) \right]},
\eeq
\noindent where $(u, v)$ denote the Fourier domain coordinates.

It is important to note that for calculating the convolution, instead of computing $s(x,y)$ in real space and taking its Fourier transform, as it is for example done in \cite{Schnars:2002}, it is better to directly calculate $S(u, v)$ using Eq. (\ref{Eq:PWFA05}), as it allows for correct sampling. The coordinates in the object plane and in the Fourier plane are sampled as explained in Appendix B. The pixel size in the Fourier domain $\Delta_F$ is given by the digital Fourier transform equation, see Eq. (\ref{Eq:B7}):
\beq\label{Eq:PWFA06}
\Delta_F =
\frac{1}{N\Delta} = \frac{1}{S},
\eeq
\noindent where $S \times S$ is the area size, $N$ denotes the number of pixels, and $\Delta$ is the pixel size in the hologram plane. In in-line holography with plane waves, the pixel size in the hologram plane $\Delta$ is equal to that in the object plane.

\newpage

\noindent The hologram simulation consists of the following steps:

\begin{tabular}{lp{15cm}}
(a)	& Calculating the Fourier transform of $t(x,y)$. All digital Fourier transforms mentioned in this work are centered, see Appendix B.\\
(b) & Simulating  $S(u,v) = \exp{\left[ - i \pi \lambda z \left( u^2 + v^2 \right) \right]}$.\\
(c) & Multiplying the results of (a) and (b).\\
(d) & Calculating the inverse Fourier transform of (c).\\
(e) & Taking the square of the absolute value of the result (d).
\end{tabular}

\vspace{0.5cm}

\noindent The reconstruction of a digital hologram recorded with plane waves is given by Eq. (\ref{Eq:PW05}) and can also can be represented as a convolution:
\beq\nonumber
U(x,y) \approx
\frac{i}{\lambda} \int\int H_0(X,Y)
\frac{\exp{\left(- ik \left|\vec{r} - \vec{R} \right| \right)}}{\left|\vec{r} - \vec{R} \right|}\;
{\rm d} X {\rm d} Y\approx
\eeq
\beq\nonumber \approx
\frac{i}{\lambda z} \int\int H_0(X,Y)
\exp{\Biggl\{- \frac{i\pi}{\lambda z}\left[\left( x - X \right)^2 + \left( y - Y \right)^2 \right]\Biggr\}}\;
{\rm d} X {\rm d} Y =
\eeq
\beq\label{Eq:PWFA07}
= H_0(x,y) \otimes s^{*}(x,y).
\eeq
\noindent The hologram reconstruction consists of the following steps:

\begin{tabular}{lp{15cm}}
(a)	& Calculating the Fourier transform of $H_0(X,Y)$.\\
(b) & Simulating $S^{*}(u,v) = \exp{\left[ i \pi \lambda z \left( u^2 + v^2 \right) \right]}$.\\
(c) & Multiplying the results of (a) and (b).\\
(d) & Calculating the inverse Fourier transform of (c). The result provides $\tilde{t}(x,y)$.
\end{tabular}

\vspace{0.2cm}
\noindent It can be shown that a convolution can also be simulated also via an inverse Fourier transforms as:
\beq\label{Eq:PWFA08}
U(x,y) =
{\rm FT} \Biggl\{{\rm FT^{-1}}\left[H_0(x,y)\right] \cdot {\rm FT^{-1}} \left[s^{*}(x,y)\right] \Biggr\},
\eeq
\noindent where
\beq\label{Eq:PWFA09}
{\rm FT^{-1}} \left[s^{*}(x,y)\right] =
\Biggl\{{\rm FT} \left[s(x,y)\right]\Biggr\}^{*} =
S^{*}(u,v),
\eeq
\noindent where $\rm FT$ and $\rm FT^{-1}$ are the Fourier transform and inverse Fourier transform respectively.

\vspace{0.2cm}
\noindent Using this approach, the hologram reconstruction consists of the following steps:

\begin{tabular}{lp{15cm}}
(a)	& Calculating the inverse Fourier transform of $H_0(X,Y)$.\\
(b) & Simulating $S^{*}(u,v) = \exp{\left[ i \pi \lambda z \left( u^2 + v^2 \right) \right]}$.\\
(c) & Multiplying the results of (a) and (b).\\
(d) & Calculating the Fourier transform of (c). The result provides $\tilde{t}(x,y)$.
\end{tabular}

\vspace{0.2cm}
\noindent At very large distances, the Fresnel condition is replaced by an even stronger Fraunhofer condition:
\beq\label{Eq:PWFA10}
z \gg \frac{\pi}{\lambda}
\left[ \left(x - X \right)^2 + \left(y - Y \right)^2 \right]_{\rm max}
\eeq
\noindent and the wave scattered by the object, given by Eq. (\ref{Eq:PWFA02}), becomes
\beq\label{Eq:PWFA11}
U_{\rm detector}(X,Y) =
- \frac{i}{\lambda z}
\exp{\left[ \frac{i}{\lambda z}\left(X^2 + Y^2 \right)\right]}
\int\int t(x,y)
\exp{\left[ - \frac{2 \pi i}{\lambda z}\left(x X +  y Y \right)\right]} \;
{\rm d} x {\rm d} y
\eeq
\noindent which is just a Fourier transform of the object transmission function $t(x, y)$. The far-field Fraunhofer condition is realized in coherent diffractive imaging \cite{Miao:1999,Latychevskaia:2012}.
\subsection{Angular Spectrum Method}

The angular spectrum method was first described by J. A. Ratcliffe \cite{Ratcliffe:1965}, and has been explained in detail by J.W. Goodman in his book \cite{Goodman:2004}. The angular spectrum method does not use any approximations. It is based on the notion, that plane wave propagation can be described by the propagation of its spectrum. The components of the scattering vector
\beq\label{Eq:PWASM01}
\vec{k} =
\frac{2\pi}{\lambda} \left( \cos{\varphi} \sin{\theta}, \sin{\varphi} \sin{\theta}, \cos{\theta}\right)
\eeq
\noindent are related to the Fourier domain coordinates $(u,v)$ as following:
\beq\nonumber
\cos{\varphi} \sin{\theta} =
\lambda u
\eeq
\beq\label{Eq:PWASM02}
\sin{\varphi} \sin{\theta} =
\lambda v
\eeq
\noindent whereby $\left( \lambda u, \lambda v\right)$ are the direction cosines of the vector $\vec{k}$, and therefore the following condition is fulfilled:
\beq\label{Eq:PWASM03}
\left(\lambda u\right)^2 + \left(\lambda v\right)^2 \le 1.
\eeq

\noindent The complex-valued exit wave $U_{\rm exit\;wave}(x,y) = t(x,y)$ is propagated to the detector plane by calculation of the following transformation \cite{Goodman:2004}:
\beq\label{Eq:PWASM04}
U_{\rm detector} (X,Y) =
{\rm FT}^{-1}
\Biggl\{
{\rm FT}\left[ t(x,y) \right]
\exp{\left[\frac{2\pi i z}{\lambda} \sqrt{1 - \left(\lambda u\right)^2 - \left(\lambda v\right)^2} \right]} \Biggr\},
\eeq
\noindent where $(u, v)$ denote the same Fourier domain coordinates as defined above. The reconstruction of the hologram is calculated by using the formula:
\beq\label{Eq:PWASM05}
U(x,y) =
{\rm FT}^{-1}
\Biggl\{
{\rm FT}\left[ H_0(X,Y) \right]
\exp{\left[ -\frac{2\pi i z}{\lambda} \sqrt{1 - \left(\lambda u\right)^2 - \left(\lambda v\right)^2} \right]}
\Biggr\}.
\eeq
\noindent The term $\displaystyle \exp{\left[ \pm \frac{2\pi i z}{\lambda} \sqrt{1 - \left(\lambda u\right)^2 - \left(\lambda v\right)^2} \right]}$ has to be simulated, and it has non-zero values for the range of $\left( \lambda u, \lambda v \right)$ constrained by Eq. (\ref{Eq:PWASM03}), which thus acts like a low-pass filter. Equation \ref{Eq:PWASM03} sets the limit for the maximal {\it possible} frequency in the Fourier domain $u_{\rm max}^{\rm max}$:
\beq\label{Eq:PWASM06}
\lambda u_{\rm max}^{\rm max} = 1.
\eeq

\noindent Taking into account Eq. (\ref{Eq:PWASM02}), we obtain: $\lambda u_{\rm max}^{\rm max} = \sin{\theta}_{\rm max}^{\rm max} = 1$, where ${\theta}_{\rm max}^{\rm max}$ is the maximal possible angle of the scattered wave. The related resolution, given by the Abbe criterion \cite{Abbe:1881,Abbe:1882} for ${\theta}_{\rm max}^{\rm max}$ amounts to:
\beq\label{Eq:PWASM07}
{\rm Resolution\; lateral} =
\frac{\lambda}{2 \sin{\theta}_{\rm max}^{\rm max}} =
\frac{\lambda}{2}.
\eeq

\noindent Thus, the condition given by Eq. (\ref{Eq:PWASM03}) relates to the classical resolution limit. Therefore, as long as imaging is done within the classical resolution limit, the condition in Eq. (\ref{Eq:PWASM03}) is always fulfilled and the wavefront propagation can be calculated by applying the angular spectrum method.	

\vspace{0.2cm}

\noindent The hologram is simulated as follows:

\begin{tabular}{lp{15cm}}
(a)	& Calculating the Fourier transform of $t(x,y)$.\\
(b) & Simulating $\exp{\left[ \frac{2\pi i z}{\lambda} \sqrt{1 - \left(\lambda u\right)^2 - \left(\lambda v\right)^2} \right]}$.\\
(c) & Multiplying the results of (a) and (b).\\
(d) & Calculating the inverse Fourier transform of (c).\\
(e) & Taking the square of the absolute value of the result (d).
\end{tabular}

\vspace{0.5cm}

\noindent The hologram reconstruction consists of the following steps:

\begin{tabular}{lp{15cm}}
(a)	& Calculating the Fourier transform of $H_0(X,Y)$.\\
(b) & Simulating $\exp{\left[ - \frac{2\pi i z}{\lambda} \sqrt{1 - \left(\lambda u\right)^2 - \left(\lambda v\right)^2} \right]}$.\\
(c) & Multiplying the results of (a) and (b).\\
(d) & Calculating the inverse Fourier transform of (c). The result provides $\tilde{t}(x,y)$.
\end{tabular}
\subsection{Resolution in In-line Holography with Plane Waves}

In general, the achievable lateral resolution in digital Gabor in-line holography is defined by \cite{Schnars:2005}:
\beq\label{Eq:PWR01}
{\rm R}_{\rm Holography} =
\frac{\lambda d}{N \Delta} =
\frac{\lambda d}{S},
\eeq
\noindent where $d$ is the distance between sample and the detector, and $S = N\Delta$ is the side length of the hologram. In practice, the resolution in in-line holography is limited by the visibility of the finest interference fringes which are formed by the interference between reference and object wave scattered at large diffraction angles. Experimentally, at least if electrons are used, the achievable resolution is often limited by the mechanical stability of the optical setup. Resolution can quantitatively be evaluated by inspecting the Fourier spectrum of a hologram \cite{Latychevskaia:2012}, similar to the resolution estimation in coherent diffractive imaging \cite{Shapiro:2005,Chapman:2006b}. Given the highest observable frequency in the Fourier spectrum $u_{\rm max}$ is detected at pixel $A$ from the center of the spectrum, its coordinate is given by:
\beq\label{Eq:PWR02}
u_{\rm max} = \Delta_F A.
\eeq

\noindent Using the relation $\sin{\theta_{\rm max}} = \lambda u_{\rm max} $, where $\theta_{\rm max}$ is the maximal detected scattering angle of the scattered wave, we obtain $\sin{\theta_{\rm max}} = \lambda \Delta_F A$. With the classical Abbe resolution criterion given by Eq. (\ref{Eq:PWASM07}) we obtain:
\beq\label{Eq:PWR03}
{\rm Resolution\;lateral} = \frac{\lambda}{2\sin{\theta}_{\rm max}} = \frac{1}{2 u_{\rm max}} = \frac{1}{2 \Delta_F A} =
\frac{S}{2 A},
\eeq
\noindent whereby we substituted $\Delta_F$ by the expression given in Eq. (\ref{Eq:PWFA06}). Thus, by estimating the position of the highest visible frequency $u_{\rm max}$ in the Fourier spectrum of a hologram, the lateral resolution intrinsic to the hologram can easily be evaluated by employing Eq. (\ref{Eq:PWR03}).

The axial resolution (in $z$-direction) can be defined as a depth of focus $\delta$. An ideal point scatterer when imaged by a diffraction-limited system will be represented as an Airy spot, with $80~\%$ of the intensity staying in the main maximum at the defocus distance \cite{Bountry:1962,Meng:1995}:
\beq\label{Eq:PWR04}
\delta =
\frac{2\lambda}{(\rm 2N.A.)^2},
\eeq
\noindent where $\rm N.A.$ is the numerical aperture of the system. This provides an estimate for the axial resolution:
\beq\label{Eq:PWR05}
{\rm Resolution\; axial} =
\frac{\lambda}{(\rm N.A.)^2}.
\eeq
\section{In-line Holography with Spherical Waves}

In this section, we describe methods of simulating and reconstructing holograms created by spherical waves, as illustrated in Fig. \ref{Fig:1}(b). This type of hologram is also called a Fresnel or Gabor hologram. The incident wave in the object plane is given by:
\beq\label{Eq:SW01}
U_{\rm incident}(x,y) =
\frac{\exp{\left( ikr \right)}}{r},
\eeq
\noindent where $\vec{r} = \left(x, y, z\right)$ and $z$ is the distance between source and object plane, as indicated in Fig. \ref{Fig:1}. The exit wave beyond the object is given by Eq. (\ref{Eq:HF03}):
\beq\label{Eq:SW02}
U_{\rm exit\; wave}(x,y) =
U_{\rm incident} (x,y) \cdot t(x,y)
= \frac{\exp{\left( ikr \right)}}{r} \cdot t(x,y).
\eeq

\noindent The propagation of the wave towards the detector is described by the Fresnel-Kirchhoff diffraction formula, see Eq. (\ref{Eq:HF04}):
\beq\label{Eq:SW03}
U_{\rm detector} (X,Y) =
-\frac{i}{\lambda} \int\int
\frac{\exp{\left( ikr \right)}}{r} \cdot
t(x,y)
\frac{\exp{\left( ik \left|\vec{r} - \vec{R} \right| \right)}}{\left|\vec{r} - \vec{R} \right|}\;
{\rm d} x {\rm d} y,
\eeq
\noindent where $\vec{r} = (x, y, z)$ is a vector pointing from the source to a point in the object, $\vec{R} = (X, Y, Z)$ is a vector pointing from the source to a point on the detector, and $\left|\vec{r} - \vec{R} \right|$ is the distance between a point in the object plane and a point in the detector plane (see Fig. \ref{Fig:1}(b)).

The reconstruction of a digital hologram recorded with spherical waves is given by Eq. (\ref{Eq:HF08}) where $R(X,Y) = \exp(ikR)/R$:
\beq\label{Eq:SW04}
U(x,y) \approx \frac{i}{\lambda}
\int\int
\frac{\exp{\left(ikR\right)}}{R} H_0(X,Y)
\frac{\exp{\left( - ik \left|\vec{r} - \vec{R} \right| \right)}}{\left|\vec{r} - \vec{R} \right|}\;
{\rm d} X {\rm d} Y.
\eeq
\subsection{Paraxial Approximation}

In the paraxial approximation, the following approximations are valid:
\beq\label{Eq:SWPA01}
r \approx z + \frac{x^2 + y^2}{2z}
\eeq
\noindent and
\beq\label{Eq:SWPA02}
\left|\vec{r} - \vec{R} \right| \approx
Z + \frac{\left(x - X \right)^2 + \left( y - Y \right)^2}{2Z}.
\eeq

\noindent They allow the following expansion of Eq. (\ref{Eq:SW03}):
\beq\nonumber
U_{\rm detector}(X,Y) =
-\frac{i}{\lambda Z z}
\exp{\left[\frac{2\pi i}{\lambda} (Z + z)\right]}
\int\int
\exp{\left[\frac{i\pi}{\lambda z} (x^2 + y^2)\right]}\;
t(x,y)\times
\eeq
\beq\label{Eq:SWPA03} \times
\exp{\Biggl\{\frac{i\pi}{\lambda Z} \left[\left(x - X\right)^2 + \left(y - Y\right)^2\right]\Biggr\}} \;
{\rm d} x {\rm d} y.
\eeq

\noindent By taking into account that $z \ll Z$, we rewrite:
\beq\nonumber
U_{\rm detector}(X,Y) =
-\frac{i}{\lambda Z z}
\exp{\left[ \frac{2\pi i}{\lambda} (Z + z)\right]}
\exp{\left[ \frac{i\pi}{\lambda Z} \left(X^2 + Y^2\right)\right]}
\int\int
\exp{\left( \frac{i\pi}{\lambda z} (x^2 + y^2)\right)}\times
\eeq
\beq\label{Eq:SWPA04}
\times t(x,y)
\exp{\left[ - \frac{2\pi i}{\lambda Z} \left(x X + y Y \right)\right]} \;
{\rm d} x {\rm d} y.
\eeq

\noindent In his original work, Gabor \cite{Gabor:1949} arrived at a similar relation, where $t(x,y)$ and $U_{\rm detector}(X,Y)$ constitute a Fourier pair, and thus $U_{\rm detector}(X,Y)$ can be obtained from $t(x,y)$ by multiplying it with a spherical phase term and taking the Fourier transform of the result, as is obvious from Eq. (\ref{Eq:SWPA04}). However, such a single Fourier transform approach is not optimal when applied to digital holograms. To design a routine for wave propagation that employs two Fourier transforms, we rewrite Eq. (\ref{Eq:SWPA04}) in the form of a convolution \cite{Latychevskaia:2012}
\beq\nonumber
U_{\rm detector}(X,Y) \approx
-\frac{i}{\lambda Z z}
\exp{\left[\frac{2\pi i}{\lambda} (Z + z)\right]}
\exp{\left[\frac{i\pi}{\lambda Z} \left(X^2 + Y^2\right)\right]}
\int\int t(x,y)\times
\eeq
\beq\label{Eq:SWPA05} \times
\exp{\Biggl\{ \frac{i \pi}{\lambda z}
\left[\left(x - X\frac{z}{Z}\right)^2 + \left(y - Y\frac{z}{Z}\right)^2 \right]
\Biggr\}} \;
{\rm d} x {\rm d} y
\eeq
\noindent of the transmission function with the Fresnel function $s(x,y)$, whereby the latter is given by Eq. (\ref{Eq:PWFA05}).

The hologram is then calculated as:
\beq\label{Eq:SWPA06}
H(X,Y) =
\left| U_{\rm detector}(X,Y)\right|^2 =
\left| t(X,Y) \otimes s(X,Y) \right|^2.
\eeq

\noindent The coordinates in the object plane respectively in the Fourier domain are sampled as explained in Appendix B. The pixel size in the Fourier domain is given by the digital Fourier transform equation, see Eq. (\ref{Eq:B7}):
\beq\label{Eq:SWPA07}
\Delta_F =
\frac{1}{N \Delta_{\rm Object}} =
\frac{1}{S_{\rm Object}},
\eeq
\noindent where $\Delta_{\rm Object} = S_{\rm Object}/N$ is the pixel size in the object plane and $S_{\rm Object} \times S_{\rm Object}$ is the object area size.

\vspace{0.2cm}
\noindent Thus, a hologram is simulated by:

\begin{tabular}{lp{15cm}}
(a)	& Calculating the Fourier transform of $t(x,y)$.\\
(b) & Simulating \newline $S(u,v) = \exp{\left[- i \pi \lambda z (u^2 + v^2) \right]}$.\\
(c) & Multiplying the results of (a) and (b).\\
(d) & Calculating the inverse Fourier transform of (c).\\
(e) & Taking the square of the absolute value of the result (d).
\end{tabular}

\noindent The size of the simulated hologram is equal to the size of the object area multiplied by the magnification factor:
\beq\label{Eq:SWPA08}
M = \frac{Z}{z}.
\eeq

\vspace{0.2cm}
\noindent The hologram is reconstructed in the reciprocal order by:

\begin{tabular}{lp{15cm}}
(a)	& Calculating the inverse Fourier transform of $H_0(X,Y)$.\\
(b) & Simulating \newline $S^{*}(u,v) = \exp{\left[i\pi\lambda z (u^2 + v^2) \right]}$.\\
(c) & Multiplying the results of (a) and (b).\\
(d) & Calculating the Fourier transform of (c). The result provides $\tilde{t}(x,y)$.
\end{tabular}

\noindent The size of the reconstructed object area is equal to the size of the hologram divided by the magnification factor $M$.
\subsection{Non-paraxial Approximation}

When the incident spherical wave extends over larger angles, the paraxial approximation is no longer valid and the field propagation based on the Fresnel-Kirchhoff diffraction formula Eq. (\ref{Eq:SW03}) must be calculated. An approach that allows the single Fourier transform integral to be transformed into the convolution integral was presented by \cite{Kreuzer:2002}. Below, we present an approach that uses propagation through the source plane \cite{Latychevskaia:2009}.

\vspace{0.2cm}
\noindent {\bf Simulation}\\
\noindent To avoid difficulties with sampling, we design a two-step routine which employing two Fourier transforms. In the first step, the wave is propagated from the object plane $\vec{r} = (x,y,z)$ to the source plane $\vec{r}_0 = (x_0,y_0,0)$. In the second step, the wave is propagated from the source plane to the detector plane \cite{Latychevskaia:2009,ZhangFucai:2004,Wang:2008}.

In the first step, with the approximation $r_0 \ll r $, we expand:
\beq\label{Eq:SWNP01}
\left|\vec{r} - \vec{r}_0 \right| \approx
r - \frac{\vec{r}\vec{r}_0}{r} + \frac{r_0^2}{2r}
\eeq
\noindent which, when substituted into Fresnel-Kirchhoff diffraction formula Eq. (\ref{Eq:SW04}), results in:
\beq\nonumber
U_0(x_0,y_0) =
\frac{i}{\lambda} \int\int
\frac{\exp{\left( ikr \right)}}{r} \cdot
t(x,y)
\frac{\exp{\left( - ik \left|\vec{r} - \vec{r}_0 \right| \right)}}{\left|\vec{r} - \vec{r}_0 \right|}\; {\rm d} x {\rm d} y \approx
\eeq
\beq\nonumber \approx
\frac{i}{\lambda}
\int\int
\frac{\exp{\left( ikr \right)}}{r} \cdot
t(x,y)
\frac{\exp{\left( - ikr \right)}}{r} \exp{\left( ik\frac{\vec{r}\vec{r}_0}{r} \right)}
\exp{\left( - ik\frac{r_0^2}{2r} \right)} \;
{\rm d} x {\rm d} y =
\eeq
\beq\label{Eq:SWNP02} =
\frac{i}{\lambda z^2}
\exp{\left[ -\frac{i\pi}{\lambda z} \left( x_0^2 + y_0^2 \right)\right]}
\int\int
t(x,y)
\exp{\left[ \frac{2\pi i}{\lambda z} (x_0 x + y_0 y) \right]}\;
{\rm d} x {\rm d} y.
\eeq
\noindent Thus, the first step consists of an inverse Fourier transform of the object transmission function $t(x,y)$ multiplied with the spherical phase term $\displaystyle \exp{\left[-\frac{i\pi}{\lambda z} (x_0^2 + y_0^2) \right]}$. The sampling in the source plane is given by the digital Fourier transform, see Eq. (\ref{Eq:B7}):
\beq\label{Eq:SWNP03}
\Delta_0 =
\frac{\lambda z}{N \Delta_{\rm Object}} = \frac{\lambda z}{S_{\rm Object}}.
\eeq

\noindent In the second step, the wavefront is propagated to the detector plane, which is described by the Fresnel-Kirchhoff diffraction formula:
\beq\label{Eq:SWNP04}
U_{\rm detector}(X,Y) =
- \frac{i}{\lambda}
\int\int U_0(x_0,y_0)
\frac{\exp{\left( ik \left|\vec{r}_0 - \vec{R} \right| \right)}}{\left|\vec{r}_0 - \vec{R} \right|}\;
{\rm d} x {\rm d} y.
\eeq

\noindent Here, the approximation $r_0\ll R$ holds and the following expansion can be applied:
\beq\label{Eq:SWNP05}
\left|\vec{r}_0 - \vec{R} \right| \approx
R - \frac{\vec{R}\vec{r}_0}{R} =
R - \vec{\kappa}\vec{r}_0,
\eeq
\noindent where we introduced the emission vector (see Appendix C):
\beq\label{Eq:SWNP06}
\vec{\kappa} = \frac{\vec{R}}{R} = \left( \frac{X}{R}, \frac{Y}{R}, \frac{Z}{R}\right) = \left(\kappa_x, \kappa_y, \kappa_z\right),
\eeq
\beq\nonumber
R = \sqrt{X^2 + Y^2 + Z^2}.
\eeq

\noindent We rewrite Eq. (\ref{Eq:SWNP04}):
\beq\nonumber
U_{\rm detector}(\kappa_x,\kappa_y) =
- \frac{i}{\lambda}
\frac{\exp{\left(ikR\right)}}{R}
\int\int U_0(x_0, y_0)
\exp{\left( - ik \vec{\kappa}\vec{r}_0 \right)}\;
{\rm d} x {\rm d} y =
\eeq
\beq\label{Eq:SWNP07} =
- \frac{i}{\lambda}
\frac{\exp{\left(ikR\right)}}{R} \int\int U_0(x_0, y_0)
\exp{\left[ - ik (x_0 \kappa_x + y_0 \kappa_y) \right]}\;
{\rm d} x_0 {\rm d} y_0.
\eeq

\noindent Thus, the second step consists of just the Fourier transform of $U_0(x_0,y_0)$. The phase factors in front of the integral vanish when the square of the absolute value is calculated and the $1/R^2$ factor cancels out after normalization of the hologram by division with the background image: $B(X,Y) = 1/R^2$. The remaining constant factor is given by $\displaystyle \frac{\Delta_0^2 \Delta_{\rm Object}^2}{\lambda^2 z^2} = \frac{1}{N^2}$.

\vspace{0.2cm}
\noindent The hologram is simulated by:

\begin{tabular}{lp{15cm}}
(a)	& Calculating the inverse Fourier transform of $t(x,y)$.\\
(b) & Simulating $\displaystyle \exp{\left[-\frac{i\pi}{\lambda z} (x_0^2 + y_0^2)\right]}$.\\
(c) & Multiplying the results of (a) and (b).\\
(d) & Calculating the Fourier transform of (c).\\
(e) & Transformation from $(\kappa_x, \kappa_y)$-coordinates to $(X,Y)$-coordinates.\\
(f)	& Taking the square of the absolute value of the result (e).\\
(g)	& Multiplication with the factor $1/N^2$.\\
\end{tabular}

\vspace{0.2cm}
\noindent {\bf Reconstruction}

\noindent The numerical reconstruction of a digital hologram consists of a multiplication of the hologram with the reference wave $R(X,Y) = e^{ikR}/R$ followed by back-propagation to the object plane given by the Fresnel-Kirchhoff diffraction formula Eq. (\ref{Eq:SW04}):
\beq\label{Eq:SWNP09}
U(x,y) \approx
\frac{i}{\lambda} \frac{\exp{\left( ikR \right)}}{R}
\int\int H_0(X,Y)
\frac{\exp{\left( - ik \left|\vec{r} - \vec{R} \right| \right)}}{\left|\vec{r} - \vec{R} \right|}\;
{\rm d} X {\rm d} Y.
\eeq

\noindent Here again, we split the reconstruction routine into two steps, which employ two Fourier transforms. In the first step the wave is propagated from the detector plane $\vec{R} = (X,Y,Z)$ to the source plane $\vec{r}_0 = (x_0,y_0,0)$. In the second step, the wave is propagated from the source plane $\vec{r}_0 = (x_0,y_0,0)$ to the object plane $\vec{r} = (x,y,z)$.

In the first step, the approximation $r_0\ll R$ is fulfilled and the expansion given by Eq. (\ref{Eq:SWNP05}) can be inserted into Eq. (\ref{Eq:SWNP09}):
\beq\nonumber
U_0(x_0,y_0) \approx
\frac{i}{\lambda}
\frac{\exp{\left( ikR \right)}}{R^2}
\int\int H_0(X,Y)
\exp{\left(- ikR\right)}
\exp{\left( ik\vec{\kappa}\vec{r}_0 \right)}\;
{\rm d} X {\rm d} Y =
\eeq
\beq\label{Eq:SWNP10} =
\frac{i}{\lambda}
\int\int H(X,Y)
\exp{\left( ik\vec{\kappa}\vec{r}_0 \right)}
J(\kappa_x,\kappa_y)\;
{\rm d} \kappa_x {\rm d} \kappa_y,
\eeq
\noindent where we took into account that  $\displaystyle H_0(X,Y) = \frac{H(X,Y)}{R^2}$ and introduced the Jacobian of the coordinate transformation:
\beq\label{Eq:SWNP11}
J(\kappa_x,\kappa_y) =
\frac{Z^2}{\left(1-\kappa_x^2 -\kappa_y^2\right)^2}.
\eeq

\noindent We rewrite Eq. (\ref{Eq:SWNP10}) as:
\beq\label{Eq:SWNP12}
U_0(x_0,y_0) =
\frac{i}{\lambda}
\int\int H(\kappa_x,\kappa_y)
\exp{\left[ik (x_0 \kappa_x + y_0 \kappa_y)\right]}
J(\kappa_x,\kappa_y)\;
{\rm d} \kappa_x {\rm d} \kappa_y
\eeq
\noindent which is simply the inverse Fourier-transform of the holographic image in  $\left(\kappa_x,\kappa_y \right)$-coordinates. The transformation of the holographic image into $\left(\kappa_x,\kappa_y \right)$-coordinates is described in Appendix C. In the second step, the field is propagated from the source plane to the object plane, which is again calculated by the Fresnel-Kirchhoff diffraction formula:
\beq\label{Eq:SWNP13}
U(x,y) =
- \frac{i}{\lambda}
\int\int U_0(x_0,y_0)
\frac{\exp{\left( ik \left|\vec{r} - \vec{r}_0 \right| \right)}}{\left|\vec{r} - \vec{r}_0 \right|}\;
{\rm d} x_0 {\rm d} y_0.
\eeq

\noindent Using the expansion given by Eq. (\ref{Eq:SWNP01}), we obtain
\beq\label{Eq:SWNP14}
U(x,y) \approx
\frac{i}{\lambda r}
\exp{\left(ikr\right)}
\int\int U_0(x_0,y_0)
\exp{\left[ - \frac{2\pi i}{\lambda z} (x_0 x + y_0 y) \right]}
\exp{\left[ \frac{i\pi}{\lambda z} (x_0^2 + y_0^2) \right]} \;
{\rm d} x_0 {\rm d} y_0,
\eeq
\noindent which is a multiplication of  with a complex spherical wave factor, followed by a Fourier transform of the result. The reconstructed exit wave includes the incident spherical wave. Thus, the result of Eq. (\ref{Eq:SWNP14}) must be divided by the incident wave to reveal the object transmission function:
\beq\label{Eq:SWNP15}
t(x,y) =
r \exp{\left(-ikr\right)} U(x,y).
\eeq

\noindent The total integral transform involved in the second step is given by
\beq\label{Eq:SWNP16}
t(x,y) \approx
\frac{i}{\lambda}
\int\int U_0(x_0,y_0)
\exp{\left[ - \frac{2\pi i}{\lambda z} (x_0 x + y_0 y) \right]}
\exp{\left[\frac{i\pi}{\lambda z} (x_0^2 + y_0^2) \right]} \;
{\rm d} x_0 {\rm d} y_0.
\eeq

\noindent When analytical integration is replaced by numerical integration, the total pre-factor turns into: $\displaystyle \frac{\Delta_0^2\Delta_{\kappa}^2}{\lambda^2} = \frac{1}{N^2}$, where $\Delta_{\kappa}$ is the pixel size in $\kappa$-space and $\Delta_0$ is the pixel size in the source plane. $\Delta_0$ is derived from $\Delta_{\kappa}$ by using Eq. (\ref{Eq:B7}):
\beq\label{Eq:SWNP17}
\Delta_0 =
\frac{\lambda}{N \Delta_{\kappa}}.
\eeq

\noindent Thus, the hologram reconstruction includes the following steps:

\begin{tabular}{lp{15cm}}
(a)	& Transforming the hologram image to $\kappa$-coordinates.\\
(b) & Calculating $J(\kappa_x, \kappa_y)$ using Eq. (\ref{Eq:SWNP11}).\\
(c) & Inverse Fourier transform of the product of (a) and (b).\\
(d) & Simulating $\displaystyle \exp{\left[\frac{i\pi}{\lambda z} (x_0^2 + y_0^2) \right]}$.\\
(e) & Multiplying the results of (c) and (d).\\
(f) & Calculating the Fourier transform of (e).\\
(g) & Multiplication with the factor $1/N^2$. The result provides $t(x,y)$.
\end{tabular}

\vspace{0.5cm}
\noindent In the algorithms for in-line holography with spherical waves, the size of a pixel in the hologram plane is equal to the size of a pixel in the object plane multiplied by the magnification factor $M$ given by Eq. (\ref{Eq:SWPA08}).
\subsection{Resolution in In-line Holography with Spherical Waves}

Similar arguments as in the discussion above on resolution concerning in-line holography with plane waves, also apply here. The practical resolution limit, intrinsic to an in-line hologram recorded with spherical waves, can be estimated from the highest frequency observed in its Fourier spectrum. The formula of the resolution, similar to Eq. (\ref{Eq:PWR03}), is given by:
\beq\label{Eq:SWR01}
{\rm Resolution\; lateral} =
\frac{S}{2A\cdot M},
\eeq
\noindent where $A$ is the pixel number at which the highest frequency in the Fourier domain is detected, $S$ is the size of the hologram, and $M$ denotes the magnification factor.
\section{Relationship between Holograms Recorded with Plane Respectively Spherical Waves}

It is worth noting that the reconstruction algorithms presented here consist of similar steps regardless of the wavefront shape: inverse Fourier transform and multiplication with the spherical phase factor followed by a Fourier transform. The spherical wave factor is given by $S^{*}(u, v)$, see Eq. (\ref{Eq:PWFA05}):
\beq\label{Eq:PWSW01}
S^{*}(u,v) =
\exp{\left[ i \pi \lambda z \left( u^2 + v^2 \right) \right]}
\eeq
\noindent in the case of plane waves and by $\displaystyle \exp{\left(\frac{i\pi}{\lambda z} (x_0^2 + y_0^2) \right)}$ in the case of spherical waves. When written in digital form, these two terms are:
\beq\nonumber
S^{*}(p, q) =
\exp{\left[ i\pi\lambda z\Delta_F^2 \left( p^2 + q^2 \right) \right]} \; {\rm and}
\eeq
\beq\label{Eq:PWSW02}
\exp{\left[\frac{i\pi}{\lambda z} \Delta_0^2\left( p^2 + q^2 \right)\right]}, \qquad p, q = 1...N,
\eeq
\noindent where $p$ and $q$ are the pixel numbers.

Substituting $\Delta_F$ and $\Delta_0$ from Eq. (\ref{Eq:PWFA06}) respectively Eq. (\ref{Eq:SWNP17}), we obtain:
\beq\nonumber
S^{*}(p, q) =
\exp{\left[\frac{i\pi\lambda z}{N^2 \Delta^2} \left(p^2 + q^2 \right)\right]} \; {\rm and}
\eeq
\beq\label{Eq:PWSW03}
\exp{\left[\frac{i\pi\lambda}{z N^2 \Delta_{\kappa}^2} \left(p^2 + q^2\right)\right]}.
\eeq

\noindent Taking into account that $\displaystyle \Delta = \frac{S_{\rm plane}}{N}$ and $\displaystyle \Delta_{\kappa} = \frac{S_{\rm spherical}}{Z N}$, where $S_{\rm plane} \times S_{\rm plane}$ and $S_{\rm spherical} \times S_{\rm spherical}$ are the related sizes of the areas in the detector plane, we obtain from Eq. (\ref{Eq:PWSW03}):
\beq\nonumber
S^{*}(p, q) =
\exp{\left[\frac{i\pi\lambda z}{S_{\rm plane}^2}\left(p^2 + q^2 \right)\right]} \; {\rm and}
\eeq
\beq\label{Eq:PWSW04}
\exp{\left[\frac{i\pi\lambda Z^2}{z S_{\rm spherical}^2}\left(p^2 + q^2\right)\right]}.
\eeq

\noindent These two terms are equal when the following equation holds:
\beq\label{Eq:PWSW05}
\frac{\lambda z}{S_{\rm plane}^2} =
\frac{\lambda Z^2}{z S_{\rm spherical}^2} =
\alpha.
\eeq

\begin{figure}[htbp]
\centerline{\includegraphics[width=12cm]{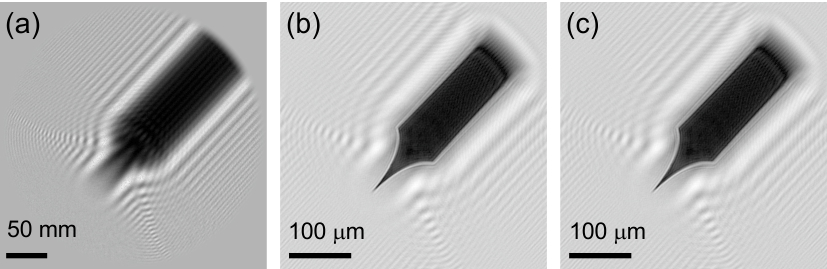}}
\caption{Optical hologram of a tungsten tip and its reconstruction. (a) Hologram recorded with $\lambda = 532$~nm laser light by the in-line Gabor scheme with the following parameters: source-to-detector distance $Z = 1060$~mm, hologram size ${S_{\rm spherical}}\times {S_{\rm spherical}} = 325 \times 325~\rm mm^2$, source-to-object distance $z = 1.4$~mm. The hologram exhibits a parameter $\alpha = 4.046\cdot 10^{-3}$. (b) Reconstructed object using the algorithm for spherical waves. The size of the reconstructed area amounts to $ 429 \times 429~\upmu{\rm m}^2$. (c) Reconstructed object assuming a planar wavefront. The hologram size is set to ${S_{\rm plane}}\times { S_{\rm plane}} = 429 \times 429~\upmu{\rm m}^2$ and the reconstruction is obtained at a hologram-to-object distance of $z = 1.4$~mm. Prior to the reconstruction, an apodization cosine-filter is applied to the edges of the normalized hologram to minimize digital Fourier transform artefacts that would otherwise arise due to a step-like intensity drop at the rim of the holographic record (see Appendix A).}\label{Fig:2}
\end{figure}

\noindent The above equation implies that a hologram recorded with a spherical wave can be reconstructed as it was recorded with plane waves, or vice versa, provided that the following relation is fulfilled:
\beq\label{Eq:PWSW06}
{S_{\rm plane}} =
\frac{z}{Z} {S_{\rm spherical}}.
\eeq
\noindent Examples of such reconstructions are shown in Fig. \ref{Fig:2}.
Moreover, Eq. (\ref{Eq:PWSW05}) implies that a uniquely defined factor $\alpha$ can be assigned to any hologram. Consequently, holograms recorded with variable wavelengths, screen sizes or source-detector distances can uniquely be reconstructed as long as $\alpha$ remains constant. This approach however is only valid for a thin object which can be assumed to be in one plane, or when the reconstruction at a certain plane within object distribution must be obtained. When reconstructing a truly three-dimensional object by obtaining a sequence of object distribution at different $z$ distances from the hologram, one must adjust the geometrical parameters at each reconstruction distance. For example, from Eq. (\ref{Eq:PWSW06}) it can be seen that the size of the reconstructed area is scaled with the distance $z$.
\section{Optimal Parameters}

During the reconstruction procedure, the inverse Fourier transform of a hologram is multiplied with the spherical phase term, which for example in the case of plane waves equals to
\beq\label{Eq:OP01}
S^{*}(u,v) =
\exp{\left[i\pi\lambda z (u^2 + v^2)\right]}.
\eeq

\noindent Such a spherical phase function is correctly simulated with $ N \times N$ pixels when it can be reduced to
\beq\label{Eq:OP02}
\exp{\left[\frac{i\pi}{N} (m^2 + n^2) \right]} \qquad m, n = 1...N,
\eeq
\noindent where $m$ and $n$ are the pixel numbers.

Taking into account the sampling $u = m \Delta_F$ and $v = n \Delta_F$, and substituting $\Delta_F$ from Eq. (\ref{Eq:PWFA06}), we obtain the following condition for correct sampling (at Nyquist or higher frequency) of the spherical phase term
\footnote{The sign in this equation has been corrected. Fresnel function $\exp{\left[\frac{i\pi}{\alpha} (m^2 + n^2) \right]}, \; m,n = 1...N$ is sampled correctly when $\alpha \ge N$. The optimal number of pixels is $N_{\rm opt} = \alpha$. When the Fresnel function $\exp{\left[i\pi\lambda z (u^2 + v^2)\right]}$ is calculated in the Fourier space where the pixel size $\Delta_F=\frac{1}{S}$, it gives $\alpha = \frac{S^2}{\lambda z}$, and the condition becomes $\alpha = \frac{S^2}{\lambda z} \geq {N}$. Here, $S$ and $z$ are parameters for the plane wave: $S$ is the sidelength of the hologram (the same as for the object area) and $z$ is the distance between the object and detector. The equation for spherical wave is $\frac{S_{\rm Object}^2}{\lambda z} \geq {N}$, where $S_{\rm Object}$ is the sidelength of the object area size and $z$ is the distance between the source and object. On the other hand, obviously, $N$ cannot be too small. The minimal $N$ should allow for correct sampling of the features of the hologram and the object.}
:
\beq\label{Eq:OP03}
\frac{S^2}{\lambda z} \geq {N}.
\eeq
\noindent This condition allows selecting optimal experimental parameters.
\section{Conclusions}

We have presented simple recipes for the numerical reconstruction of in-line holograms recorded with plane and spherical waves. These methods are wavelength-independent and can thus be applied to holograms recorded with any kind of radiation exhibiting wave nature. Moreover, reconstructions of both absorbing as well as phase shifting properties of objects can be achieved. We also demonstrated that any digital hologram can be assigned a uniquely defined parameter which defines its digital reconstruction.\\
\vspace{0.2cm}

\noindent {\bf Acknowledgments}

\noindent Financial support by the Swiss National Science Foundation and the University of Zurich are gratefully acknowledged.
\newpage
\appendix
\section{Apodization Cosine Filter}
\setcounter{equation}{0}
\numberwithin{equation}{section}

\begin{figure}[htbp]
\centerline{\includegraphics[width=0.5\columnwidth]{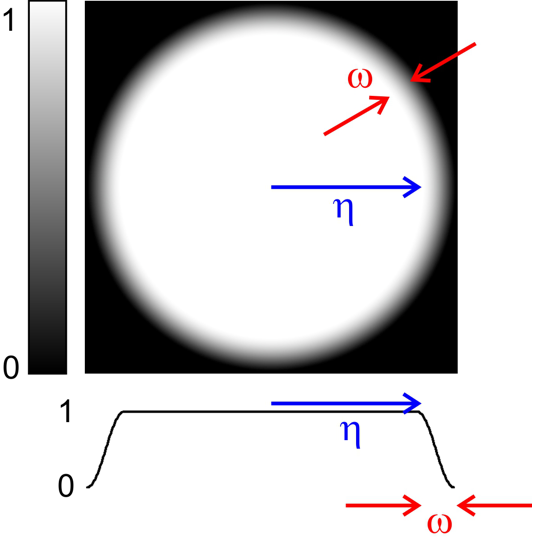}}
\caption{Intensity distribution of the apodization cosine filter and its amplitude profile through the center of the image.}\label{Fig:3}
\end{figure}

Before reconstruction, the normalized hologram given by Eq. (\ref{Eq:HF07}) is multiplied with an apodization cosine filter to smooth the intensity at the hologram edges down to zero and thus minimize the effects on the edges of the hologram due to digital Fourier transform. The cosine filter is mathematically described by the function
\beq\label{Eq:A1}
C(\rho) =
\begin{cases}
\cos^2{\left[ \frac{\pi}{2\omega} (\rho - \eta) \right]}, & \eta < \rho < \eta+\omega\\
1, & 0 < \rho < \eta \\
0, & \rho > \eta + \omega,
\end{cases}
\eeq
\noindent where $\rho = \sqrt{X^2 + Y^2}$ and $C(\rho)$ distribution is shown in Fig. \ref{Fig:3}.\footnote{this equation appears in wrong format in the online version of the paper (Applied Optics 54(13), 2424 -- 2434 (2015))}.
\newpage
\section{From Analytical to Fast Fourier Transform (FFT)}
\numberwithin{equation}{section}

The analytical Fourier transform connecting two domains is given by:
\beq\label{Eq:B1}
U_2(u,v) = \int\int U_1(x,y) \exp{\left[-2\pi i (xu + yv)\right]}\;
{\rm d}x {\rm d}y.
\eeq

\noindent The fast Fourier transform is calculated by using the following expression:
\beq\nonumber
U_2(p, q) =
{\rm FFT} \left(U_1 (p, q)\right) =
\eeq
\beq\label{Eq:B2} =
\sum\limits_{m, n=1}^{N} U_1(m, n)
\exp{\left[-\frac{2\pi i}{N} (mp + nq)\right]},
\eeq
\noindent where $m$, $n$, $p$ and $q$ are the pixel numbers.

For a numerical calculation of the analytical transform by means of FFT, the distribution $U_1(x,y)$ has to be sampled in $(x,y)$ coordinates. Assuming that the center of the coordinate system $(x,y)$ is at the center of the distribution $U_1(x,y)$, the sampling is done at the points
\beq\nonumber
x =
\left(m - \frac{N}{2}\right)\Delta_1,\; {m = 1...N},\;{\rm and}
\eeq
\beq\label{Eq:B3}
y =
\left(n - \frac{N}{2}\right)\Delta_1,\; {n = 1...N}
\eeq
\noindent and in the Fourier domain:
\beq\nonumber
u =
\left(p - \frac{N}{2}\right)\Delta_2,\; {p = 1...N},\;{\rm and}
\eeq
\beq\label{Eq:B4}
v =
\left(q - \frac{N}{2}\right)\Delta_2,\; {q = 1...N}.
\eeq

\noindent Here $\Delta_1$ and $\Delta_2$ are the pixel sizes in the two domains respectively, given by:
\beq\label{Eq:B5}
\Delta_1 =
\frac{S_1}{N}\qquad{\rm and}\qquad \Delta_2 = \frac{S_2}{N},
\eeq
\noindent where $S_1 \times S_1$ and $S_2 \times S_2$ are the sizes of the areas in the related domains.

\noindent For the phase term in Eq. (\ref{Eq:B1}) we obtain:
\beq\label{Eq:B6}
2\pi (xu + yv) =
2\pi \Delta_1 \Delta_2 \left[mp + nq - \frac{N}{2}(m + p + n + q) + \frac{N^2}{2}\right],
\eeq
\noindent which equals the phase term in Eq. (\ref{Eq:B2}) provided the following condition is fulfilled:
\beq\label{Eq:B7}
\Delta_1 \Delta_2 = \frac{1}{N}.
\eeq

\noindent By substituting Eq. (\ref{Eq:B6}) into Eq. (\ref{Eq:B1}) and skipping the last term in Eq. (\ref{Eq:B6}), since it does not contribute to the phase, we obtain the formula for calculating the centered Fourier transform:
\beq\nonumber
U_2(p, q) =
\exp{\left( i\pi (p + q) \right)}
\sum\limits_{m, n = 1}^{N} U_1(m, n)
\exp{\left[-\frac{2\pi i}{N} (mp + nq)\right]}
\exp{\left[ i\pi (m + n) \right]}=
\eeq
\beq\label{Eq:B8} =
\exp{\left[ i\pi (p + q) \right]}
{\rm FFT} \Biggl\{ U_1(m, n) \exp{\left[ i\pi (m + n) \right]} \Biggr\}.
\eeq

\noindent The inverse Fourier transform is calculated in a similar manner:
\beq\label{Eq:B9}
U_2(p, q) =
\exp{\left[- i\pi (p + q) \right]}
{\rm FFT}^{-1} \Biggl\{ U_1(m, n) \exp{\left[- i\pi (m + n) \right]}\Biggr\}.
\eeq
\newpage
\section{$\kappa$-coordinates}
\numberwithin{equation}{section}

Considering the emission vector coordinates given by Eq. (\ref{Eq:SWNP06}), we obtain the maximal value for $\kappa_x$:
\beq\label{Eq:C1}
\kappa_{x,{\rm max}} =
\frac{S/2}{\sqrt{(S/2)^2 + Z^2}}
\eeq
\noindent where $S \times S$ represents the detector size which provides the pixel size in $\kappa$-coordinates:
\beq\label{Eq:C2}
\Delta_\kappa =
\frac{2 \kappa_{x,{\rm max}}}{N}.
\eeq

\noindent The transformation of the hologram from $(X, Y)$ coordinates to $(\kappa_x, \kappa_y)$ coordinates includes the following steps. The arrays of $(\kappa_x,\kappa_y)$ are created as follows:
\beq\nonumber
\kappa_x =
\left(m - \frac{N}{2}\right)\Delta_\kappa,\; {m = 1...N},\;{\rm and}
\eeq
\beq\label{Eq:C3}
\kappa_y =
\left(n - \frac{N}{2}\right)\Delta_\kappa,\; {n = 1...N}.
\eeq

\noindent Next, the value at each $(\kappa_x, \kappa_y)$ pixel is assigned the value at a pixel $(X, Y)$ in the hologram $H(X, Y)$, where:
\beq\label{Eq:C4}
X = \frac{Z\kappa_x}{\sqrt{1 - \kappa_x^2 - \kappa_y^2}},\qquad
Y = \frac{Z\kappa_y}{\sqrt{1 - \kappa_x^2 - \kappa_y^2}}.
\eeq

\begin{figure}[htbp]
\centerline{\includegraphics[width=8cm]{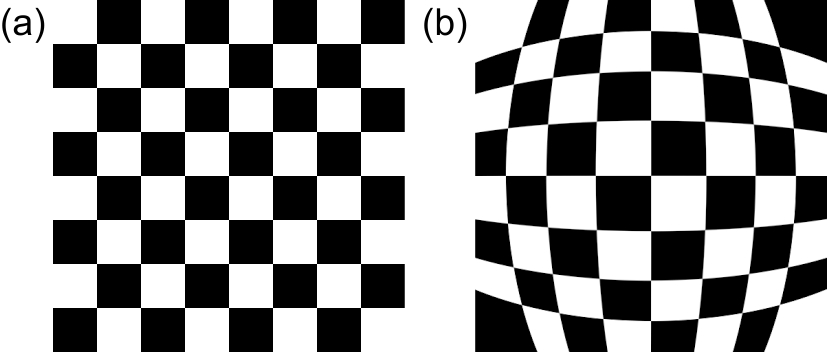}}
\caption{Example of the appearance of an image when it is transformed from $(X, Y)$ to $(\kappa_x, \kappa_y)$ coordinates. (a) Chessboard image in $(X, Y)$ coordinates. The image is assumed to have a size of $0.8 \times 0.8~\rm m^2$ on a detector which is $0.5$~m away from the source. These parameters correspond to a numerical aperture $= 0.625$. (b) Chessboard image in $(\kappa_x, \kappa_y)$ coordinates.}\label{Fig:4}
\end{figure}

\noindent The transformation depends only on the numerical aperture of the system, that is, on the source-to-detector distance and the detector size. The transformation resembles visually the "fish eye" effect, as illustrated in Fig. \ref{Fig:4}. The reverse transformation from $(\kappa_x, \kappa_y)$ coordinates to $(X, Y)$ is done in a similar manner.

\end{document}